# Network Performance Rank: An Approach for Comparison of Complex Networks

Zeynab Bahrami Bidoni [1], Roy George [2]
*Department of Computer and Information Systems*
*Clark Atlanta University*
*Atlanta, GA*
*z.bahrami62@gmail.com* [1]
*rgeorge@cau.edu* [2]

**Abstract**

Researchers have typically concentrated on analyzing what happens internally in a complex network and using this to distinguish between nodes. However, there has been less effort towards comparing between different networks. In this paper, we proposed a novel approach to rank alternative complex networks based on their performances. We consider this as a ranking problem in decision analysis based on occurring positive/negative frequent events as criteria, and using the TOPSIS method to rank alternatives. In order to assign a score to the networks for each criterion, a statistical method that estimates the expected value of positive/negative frequent events on a random node is presented. The proposed technique is efficient in terms of algorithm complexity and is capable of discriminating events occurring between important nodes over those between less significant nodes. The experiments, conducted on several synthetic networks, demonstrate the feasibility and applicability of the ranking methodology.

## 1. Introduction

In current years, researchers have mostly focused on the internals of complex networks developing techniques such as detecting communities [1-12], ranking nodes [13-18], finding outliers [19-21], and etc. There has been less attention given towards the performance comparisons between different networks. This problem manifests itself in different domains including Computer, Telecommunication, Electrical Circuit, Supply Chain, Social networks etc., where, there is a need to evaluate different network architectures, equipment, protocols etc. with the constraint, that it is not possible to replicate the exact same scenarios in each case. In this study, we assume this objective as a Ranking problem in Multi-Criteria Decision Making (MCDM) [22-26] field based on occurring positive/negative frequent events as the criteria.

Since any event occurs between two nodes of a network, and the nodes could not considered as independent variables, statistical analysis to compute the probability of failed/successful occurrence between random nodes throughout the network would be very difficult. This paper proposes a novel approach to approximate variance of all type of event per networks, which is used to estimate the expected values of the events between two random nodes.

This research is funded in part by the Army Research Laboratory under Grant No: W911NF-12-2-0067 and Army Research Office under Grant Number W911NF-11-1-0168. Any opinions, findings, conclusions or recommendations expressed here are those of the author(s) and do not necessarily reflect the views of the sponsor.

The contributions of this paper are as follows: (1) Defining the networks performance comparison problem as a Multi-Criteria Decision Making (MCDM) ranking problem, (2) Developing an approach to compute the diversity of density (DOD) of events in networks to evaluate the variance where the events happening between important nodes are positively discriminated over events between less significant nodes. (3) Approximating the probability distribution and the expected value of occurrences on a random node for scoring each network per criteria.

The rest of the paper is organized as follows. Section 2 presents the general framework and all approaches needed for ranking alternative networks. Section 3 provides the experimental results. Section 4 offers concluding remarks.

## 2. proposed approach to compare between networks

### 2.1 General framework

In order to compare between networks performance, we consider this issue as a ranking problem in MCDM. A MCDM problem can be concisely expressed in matrix format as

|       | $C_1$    | $C_2$    | $\ldots$ | $C_n$    |
|-------|----------|----------|----------|----------|
| $A_1$ | $x_{11}$ | $x_{12}$ | $\ldots$ | $x_{1n}$ |
| $A_2$ | $x_{21}$ | $x_{22}$ | $\ldots$ | $x_{2n}$ |
| $A_m$ | $x_{m1}$ | $x_{m2}$ | $\ldots$ | $x_{mn}$ |

$$W = [w_1, w_2, \ldots, w_n]$$

Figure 1. A decision matrix in MCDM problem model

Where $A_1, A_2, \ldots, A_m$ are possible alternative networks among which have to rank, $c_1, c_2, \ldots, c_n$ are criteria with which alternative performance are measured, $x_{ij}$ is the score of alternative $A_i$ with respect to criterion $c_j$, $w_j$ is the weight of criterion $c_j$.

Various attributes can be selected as criteria but, here, we focus on positive/negative frequent events which may occur between nodes during a given big enough period of time, and effect on network's performance. For instance, the re-transaction or any failed operation between nodes is the negative event in net which decrease network's efficiency. On the contrary, the more probability density of successful and positive occurrence is the more efficiency will be in the network.

So, for establishing the decision matrix follow steps needed:

a) Select the collection of criteria.
b) Scoring networks on criteria.

After that, using TOPSIS method which explained in section 2-3 to rank the alternative networks.

*2.2 Scoring networks on criteria*

With the purpose of scoring networks on criteria, we proposed a new density based approach which compute the global DOD of the given positive/negative frequent event (criteria) for each networks. Gaussian distribution is employed based on the average number of events per unit as the mean parameter and the approximated DOD is used as the variance parameter to estimate the expected value of the event frequency between two random nodes per network during given big enough time periods.

In order to compute the global DOD on given criterion, we used the modified ''Correlation Density Rank'' Method [27] which finds probability density distribution of the related frequent event on all nodes, and then we utilize the Renyi entropy [28] to realize the global unpredictability or diversity of these densities on whole network.

2) *Correlation Density Rank*

We use the Correlation Density Rank (CDR), [27] which finds more frequent and influential Randomized shortest Path (RSP). The CDR considers the distance between nodes as punishment and is used to compute probability density of nodes. Hence, there will be a larger traffic amongst shortest path of nodes, if the distance becomes smaller. Therefore, the objective is to minimize punishment so that a node with high value of density probability to have a higher rank.

Moreover, the more popular nodes are the more linkages other nodes tend to have to them or are linked to by them. The proposed algorithm is analogous to the weighted PageRank algorithm [29, 30], assigning larger rank values to more important (popular) nodes instead of dividing the rank value of a node evenly among its out-link nodes. We assign each out-link node a value proportional to its popularity (its number of in-links and out-links). The popularity from the number of in-links and out-links is recorded as $W_{ij}^{in}$ and $W_{ij}^{out}$, respectively.

$W_{ij}^{in}$ is the weight of link between node $n_i$ and $n_j$ calculated based on the number of in-links of node $n_j$ and the number of in-links of all reference nodes of node $n_i$.

$$W_{ij}^{in} = \frac{I_{n_j}}{\sum_{p \in R(n_i)} I_p} \quad (1)$$

Where $I_{n_j}$ and $I_p$ represent the number of frequency in-links of node $n_j$ and node $p$, respectively. R($n_i$) denotes the reference node list of node $n_i$.

$W_{ij}^{out}$ is the weight of link between node $n_i$ and $n_j$ calculated based on the number of frequency out-links of node $n_j$ and the number of out-links of all reference nodes of node $n_i$.

$$W_{ij}^{out} = \frac{O_{n_j}}{\sum_{p \in R(n_i)} O_p} \quad (2)$$

Where $O_{n_j}$ and $O_p$ represent the number of out-links of node $n_j$ and node p, respectively. R($n_i$) denotes the reference node list of node $n_i$. These equations has two exceptions, first, if node $n_j$ is a dead-end (which may be easily determined from the frequency matrix), we let $W_{ij}^{out} = \varepsilon$ that $\varepsilon$ is a very small number less than 1. Second, $W_{ij}^{out}, W_{ij}^{in} = 1$, that means R($n_i$)={$n_j$} we add $\varepsilon$ to sum of the reference nodes' frequency out/in-link.

An algorithm for calculating the probability density of related frequent event for all members in a complex network is described as follows.

Algorithm 1. Correlation Density Rank (CDR):

Input: social network G

Out: vector of probability density distribution CDR

1. Initialize cost distance matrix C

$$C[i,j] = \log \frac{(1-\exp(-\gamma f_{ij}))}{(1-w_{ij}^{in} w_{ij}^{out})} \quad (3)$$

(The logarithm of $(1-\exp(-\gamma f_{ij}))$ based on $(1-w_{ij}^{in} w_{ij}^{out})$)

2. Finding the matrix of RSP dissimilarities by employ the algorithm of [29]:

{

$$W \leftarrow P^{ref} \circ \exp(-\beta C) \quad (4)$$

$$Z \leftarrow (I - W)^{-1} \quad (5)$$

(Note that $(I-W)^{-1} \approx I + W + W^2 + W^3 + ...$)

$$S \leftarrow (Z(C \circ W)Z) \div (Z + \varepsilon) \quad (6)$$

$$\tilde{C} \leftarrow S - ed_s^T \quad (7)$$

$$\Delta^{RSP} \leftarrow 0.5(\tilde{C} + \tilde{C}^T) \quad (8)$$

}

3. $M \leftarrow$ Normalize matrix $\Delta^{RSP}$ on columns

4. For each node $nj$ ($1 \le j \le k$) compute inverse of the entropy [31] of related column from matrix M ($\sigma j$ is the jth kernel scale parameter which describes the influence of a node $nj$ within its Neighborhood. we optimize σ for each node to make the density values the most different):

$$ej \leftarrow -\frac{1}{Lnk}\sum_{i=1}^{k} M_{ij} Ln(M_{ij}) \quad (9)$$

$$\sigma j \leftarrow \frac{1}{e_j} \quad (10)$$

5. Calculate the density function which results from a Gauss Influence function [32] (it sorts all the nodes in descending order according to their CDR values)

$$cdr_i \leftarrow \sum_{j=1}^{k} \exp\left(-\frac{\left(\Delta_{ij}^{RSP}\right)^2}{2\sigma_j^2}\right) \quad (11)$$

6. Normalize Correlation Density Rank vector (we can sort all the nodes in descending order according to their CDR values):

$$CDR_i \leftarrow cdr_i \Big/ \sum_{i=1}^{k} cdr_i \quad (12)$$

7. Return CDR.

Where $f_{ij}$ is the number of frequency from node $n_i$ to node $n_j$. if $f_{ij} = 0$, let $C[i,j] = \infty$ or a very big number. $P^{ref}$ is the transition probability matrix that $P_{ij}^{ref}$ is equal to the rate of $f_{ij}$ divided by sum of frequency between node $n_i$ and all its references nodes. k is the number of members in social network (or nodes on G).

The parameters $\gamma$ and $\beta$ are input values determined by user. $\gamma$ controls the effect of frequency on the cost function which restrict cost ratio with respect to our defined infinite constant. $\beta$ is the influence of the cost on the walker's selection of a path, and is equal to inverse of temperature at Helmholtz free energy in thermodynamical system [33].

Also, in step 2, $d_s = diag(S)$ is the vector of diagonal elements of $S$, and $e$ is the identity matrix. Note that A ∘ B and A ÷ B are elementwise product and division, respectively.

For calculating step 2, we use the easier way of computing the matrix Z [34]. The values of $CDR_i$ ($1 \le i \le k$) indicate the final normalized density rank of members in the complex network which are considered as probability density distribution on nodes.

3) *Measure of the global unpredictability/ DOD for each criterion per network*

The Shannon entropy is a measurement of system uncertainty, unpredictability, diversity and randomness [31] and has been used in statistics and information theory to develop measures of the information content [35]. The larger the Shannon entropy is, the more uncertainty/unpredictability/randomness and less diversity of the system will be. Also, Shannon entropy is the classical measure of information content and is defined for an n-dimensional probability density (PD) distribution P(x) as:

$$H(P) = \int_{-\infty}^{\infty} P(x) \log P(x) dx \quad (13)$$

Since several time frequency representations can achieve negative values the use of the more classical Shannon information as a measure of complexity is prohibited (due to the presence of the logarithm within the integral in below) and some authors [28, 36-38] have proposed the use of a relaxed measure of entropy known as the Renyi entropy of order $\alpha$:

$$H_\alpha^R(P) = \frac{1}{1-\alpha} \log \frac{\int P^\alpha(x) dx}{\int P(x) dx} \quad (14)$$

Following Baraniuk, the passage from the Shannon entropy H to the class of Renyi entropies $H_\alpha^R$ involves only the relaxation of the mean value property from an arithmetic to an exponential mean and thus in practice $H_\alpha^R$ behaves much like H. The Shannon entropy can be recovered as $\lim_{a \to 1} H_\alpha^R(P) = H(P)$.
(15)

So, in order to measure of the global DOD/unpredictibility for each network, we can employ the CDR vector as the probability density distribution on nodes in Renyi entropy formulate. Thus, for scoring each network on each criteria, we compute the follow measure:

$$H_k^{c_l} = \frac{1}{1-\alpha} \log_2 \left( \frac{\sum_{i=1}^{N_k} CDR_i^\alpha}{\sum_{i=1}^{N_k} CDR_i} \right) \quad (16)$$

Where $H_k^{c_l}$ the unpredictability of network number $k$ on the event related to criterion $c_l$ and $N_k$ is the number of nodes in network number $K$. Also, CDR vector is related to given network and event, and $\alpha$ is the order of Renyi entropy order that we can consider 3.

If the density value of each node in complex network is the same, then the uncertainty of the original density distribution is the greatest. On the contrary, while the density value of each node in complex network is very asymmetrical, then the uncertainty of the original density distribution is the smallest. Thus, the inverse of uncertainty/unpredictability would be a good measure for the DOD through network.

4) *Estimate the expected value rate of each event per each networks*

As mentioned above, the Renyi entropy can reflect the difference of nodes' density value. The more different the density values are, the smaller the Renyi entropy is. So, we can consider inverse of Reyni entropy's result time the mean as a measure of the frequent event variance through nodes.

The normal (or Gaussian) distribution is a very commonly occurring continuous probability distribution—a function that tells the probability that an observation in some context will fall between any two real numbers. Normal distributions are extremely important in statistics and are often used in the natural and social sciences for real-valued random variables whose distributions are not known [39]. Furthermore, considering a Guassian distribution with below mean and variance parameters, can help us to better understanding about probability distribution of given frequent event through $k$th network.

$$\mu_k^{c_l} = \frac{\sum_{i=1}^{N_k}\sum_{j=1}^{N_k} f_{ij}}{N_k} \quad , \quad \delta_k^{c_l} = \frac{\mu_k^{c_l}}{H_k^{c_l}} \qquad (17)$$

$$F_k\left(x, \mu_k^{c_l}, \delta_k^{c_l}\right) = \frac{1}{\delta_k^{c_l}\sqrt{2\pi}} e^{-\frac{\left(x - \mu_k^{c_l}\right)^2}{2(\delta_k^{c_l})^2}} \qquad x \geq 0 \qquad (18)$$

For instance, probability of not occur the given event on a random node in $k$th network would be result of below equation:

$$p(x = 0) = \frac{e^{-\frac{\left(\mu_k^{c_l}\right)^2}{2(\delta_k^{c_l})^2}}}{\delta_k^{c_l}\sqrt{2\pi}} \qquad (19)$$

Moreover, the expected value of event on two random nodes in networks' Gaussian distribution is a good measure for scoring networks on related criteria. Thus, we have

$$S_k^{c_l} = \int_0^{+\infty} x F_k\left(x, \mu_k^{c_l}, \delta_k^{c_l}\right) dx = \int_0^{+\infty} \frac{x}{\delta_k^{c_l}\sqrt{2\pi}} e^{-\frac{\left(x - \mu_k^{c_l}\right)^2}{2(\delta_k^{c_l})^2}} dx \qquad (20)$$

Where $S_k^{c_l}$ is the score of $k$th network on criterion $c_l$. After using this approach for all networks and scoring them on all criteria, Decision Matrix is constructed to apply TOPSIS method for ranking networks' performance. Note that before start TOPSIS's steps normalize Decision matrix on rows to have the ratio of expected value for each networks. Because, networks may have different amounts of global expected values of all event types on a random node. So, for fair compare between them, the ratio of expected value is a better measure to evaluate.

## 2.3 TOPSIS algorithm

Many ranking methods have been proposed to solve the multiple criteria decision making (MCDM) problems, etc. One of the well-known ranking methods for MCDM, named the technique for order preference by similarity to ideal solution (TOPSIS) [40-45], is firstly proposed by Hwang and Yoon [46]. The logic of the TOPSIS approach is to define the ideal and anti-ideal solutions [43], which are based on the concept of relative closeness in compliance with the shorter (longer) the distance of alternative i to ideal (anti-ideal), the higher the priority can be ranked [47]. The procedure of TOPSIS can be expressed in a series of steps:

(1) Calculate the normalized decision matrix on column. The normalized value $n_{ij}$ is calculated as

$$n_{ij} = x_{ij} \Big/ \sqrt{\sum_{i=1}^{m} x_{ij}^2} \qquad j = 1,...,n, \quad i = 1,...,m. \qquad (21)$$

(2) Calculate the weighted normalized decision matrix. In this paper, the weights of objective criteria, using the entropy weighting method [47], will be applied to the generalized algorithm. The weighted normalized value $v_{ij}$ is calculated as

$$v_{ij} = w_j n_{ij}, \qquad j = 1,...,n, \quad i = 1,...,m, \qquad (22)$$

Where $wj$ is the weight of the *i*th attribute or criterion, and $\sum_{i=1}^{n} wj = 1$.

(3) Determine the positive ideal and negative ideal solution.

$$A^+ = \{v_1^+,...,v_n^+\} = \left\{\left(\max_i v_{ij} \Big| j \in I\right), \left(\min_i v_{ij} \Big| j \in J\right)\right\},$$

$$A^- = \{v_1^-,...,v_n^-\} = \left\{\left(\min_i v_{ij} \Big| j \in I\right), \left(\max_i v_{ij} \Big| j \in J\right)\right\},$$

$$(23)$$

Where *I* is associated with benefit criteria, and *J* is associated with cost criteria.

(4) Calculate the separation measures, using the m-dimensional Euclidean distance. The separation of each alternative from the ideal solution is given as

$$d_i^+ = \left\{\sum_{j=1}^n \left(v_{ij} - v_j^+\right)^2\right\}^{\frac{1}{2}}, \qquad i = 1,2,...,m. \qquad (24)$$

Similarly, the separation from the negative ideal solution is given as

$$d_i^- = \left\{ \sum_{j=1}^{n} \left(v_{ij} - v_j^-\right)^2 \right\}^{\frac{1}{2}}, \quad i = 1, 2, ..., m. \qquad (25)$$

(5) Calculate the relative closeness to the ideal solution. The relative closeness of the alternative $A_j$ with respect to $A^+$ is defined as

$$R_i = d_i^- / \left(d_i^+ + d_i^-\right), \quad i = 1, ..., m. \qquad (26)$$

Since $d_j^- \geq 0$ and $d_j^+ \geq 0$, then, clearly, $R_j \in [0,1]$.

(6) Rank the preference order. For ranking networks using this index, we can rank networks' the relative closeness value in decreasing order.

The basic principle of the TOPSIS method is that the chosen alternative should have the ''shortest distance'' from the positive ideal solution and the ''farthest distance'' from the negative ideal solution. The TOPSIS method introduces two ''reference'' points, but it does not consider the relative importance of the distances from these points.

## 3. EXPERIMENTS

In order to implement our approach, we designed four different synthetic architectures of computer network with recording the successful and failed type of frequencies as positive and negative events respectively, during sample time period which are shown on Figure 2. Networks' data can show obviously our method behavior on different situation. For instance, Network A and B have same number of successful and failed events, But the DOD of failed events in network A is higher than network B so that it seems node 8 in network A has a critical problem and probability of happening failed events between node 8 and any other nodes is high.

After employing correlation density rank and Renyi entropy we have found global unpredictability and variance of events per networks, which mentioned in Table 1. within the other general information.

Table 1. Network properties and their unpredictability, mean and variance results by proposed method.

| Network Name | Number of nodes | Successful events | | | | Fail events | | | |
|---|---|---|---|---|---|---|---|---|---|
| | | Number of frequency | Unpredictability | Mean | Variance | Number of frequency | Unpredictability | Mean | Variance |
| Network A | 10 | 100 | 2.6610 | 10 | 3.757 | 25 | 0.01 | 2.5 | 250 |
| Network B | 12 | 100 | 3.0501 | 8.333 | 2.732 | 25 | 1.7552 | 2.083 | 1.186 |
| Network C | 8 | 60 | 2.0871 | 7.5 | 3.5935 | 30 | 2.1121 | 3.75 | 1.7754 |
| Network D | 9 | 60 | 1.4401 | 6.6667 | 4.62933 | 10 | 0.7935 | 1.1111 | 1.4002 |

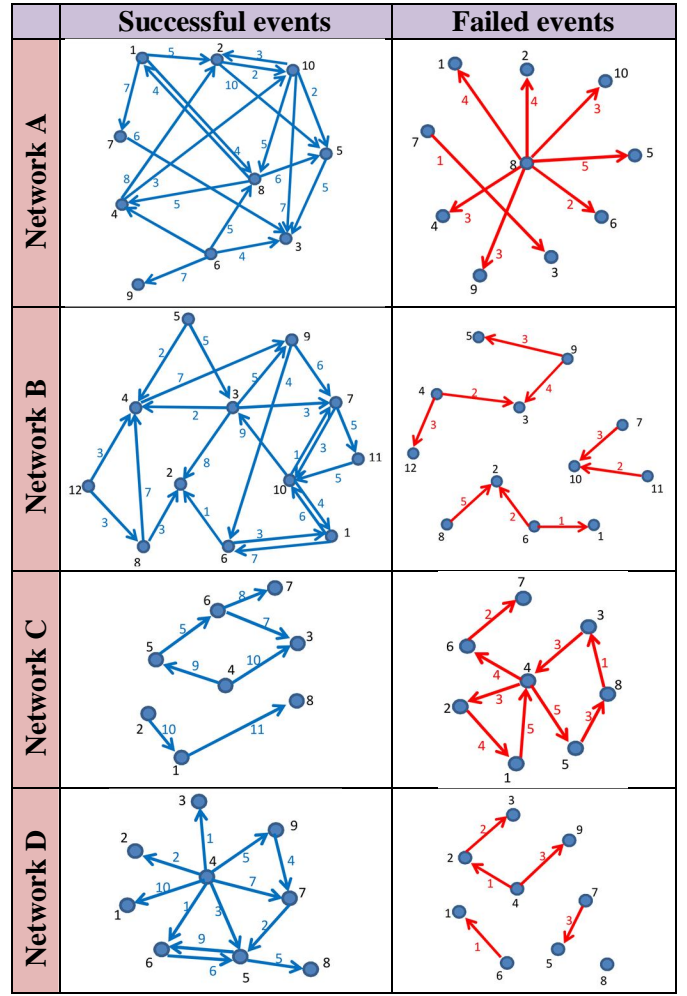

Figure 2. Four synthetic architectures of computer network with their successful and failed frequencies during sample time period.

By having the estimated mean and variance parameters, we can consider networks' probability distributions on both successful and failed events (Figure 3 and Figure 4). For example in Figure 4, the junction of probability distribution curves and Y axis indicate the probability of no occurring failed event on a random node on related network that in this case network D, B, C and A respectively have descending order of the probability of no occurring failed event on a random node.

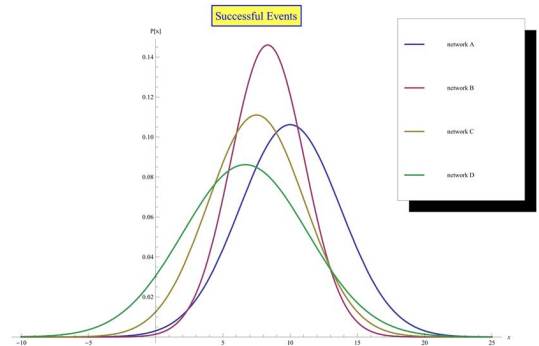

Figure 3. Estimated probability distributions for networks about successful type of events.

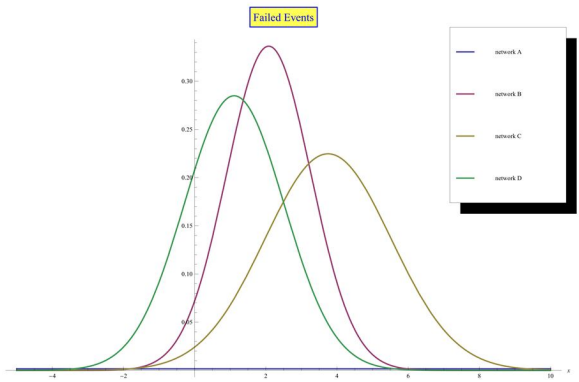

Figure 4. Estimated probability distributions for networks about failed type of events.

In next stage, decision matrix constructed as shown in Table 2 and then using Entropy weighting Method, weights of criteria were computed (Table 3).

Table 2. Decision matrix.

| Network Name | Criterion 1 Expected value of successful event | Criterion 2 Expected value of failed event |
|---|---|---|
| Network A | 10.0045 | 100.991 |
| Network B | 8.3387 | 2.10188 |
| Network C | 7.52409 | 3.7611 |
| Network D | 6.822 | 1.28134 |

Table 3. Rsults of weighting criteria by Entropy weighting method.

| Entropy weighting method | Criterion 1 successful event | Criterion 2 failed event |
|---|---|---|
| weight | 0.6425 | 0.3575 |

Finally, the Topsis Method helped us to rank networks based on these two criteria (Table 4).

Table 4. Results of ranking by TOPSIS

| Network Name | Network A | Network B | Network C | Network D |
|---|---|---|---|---|
| Rank value by Topsis method | 0 | 0.942537 | 0.766967 | 1 |

As expected, rank values by TOPSIS method displayed that network D is the best one and networks B, C and A, respectively, have smaller ranks on descending order.

## 4. Conclusions

Ranking complex networks has a broad range of applications, such as Computer/Corporate/Campus Area Network (CAN), Telecommunication Network, Electrical Circuit Network, Social Network, Supply Chains, Financial networks and etc. In this paper, we present our research effort in comparing between complex networks from their positive/negative frequency data to obtain a ranking of them. The proposed method is composed of three main parts: (1) an approach for estimating the DOD of event frequencies through network; (2) a static framework to explore the expected value of each type of events frequency on a random node per network which is considered as the score of network on related criterion; and, (3) construct the decision matrix and employ the well-known TOPSIS method to rank alternative networks. These algorithms were applied to several synthetic datasets, and produce good results. The experiments show that the framework can well present the rank order of networks.